\documentclass[prl,preprint,superscriptaddress,aps]{revtex4}
\usepackage{amsmath}
\usepackage{graphicx}
\usepackage{epsfig}
\usepackage{psfrag}
\usepackage{mathdots}
\usepackage{comment}
\usepackage{natbib}
\usepackage{color}
\newlength{\upit}\upit=0.1truein

\begin{document}

\title{Kondo insulator SmB$_6$ under strain: surface dominated conduction near room temperature}

\author{A. Stern}
\affiliation{Department of Physics and Astronomy, University of California, Irvine, California 92697, USA}

\author{M. Dzero}
\affiliation{Department of Physics, Kent State University, Kent, OH 44242, USA}

\author{V. M. Galitski}
\affiliation{Condensed Matter Theory Center, Department of Physics, University of Maryland, College Park, Maryland 20742, USA}

\author{Z. Fisk}
\affiliation{Department of Physics and Astronomy, University of California, Irvine, California 92697, USA}

\author{J. Xia}
\affiliation{Department of Physics and Astronomy, University of California, Irvine, California 92697, USA}

\begin{abstract}
SmB$_6$ is a strongly correlated mixed-valence Kondo insulator with a newly discovered surface state, proposed to be of non-trivial topological origin.  However, the surface state dominates electrical conduction only below $T\textsuperscript{*} \approx$ 4 $\textit{K}$ limiting its scientific investigation and device application. Here, we report the enhancement of $T\textsuperscript{*}$ in SmB$_6$ under the application of tensile strain. With 0.7\% tensile strain we report surface dominated conduction at up to a temperature of 240 $\textit{K}$, persisting even after the strain has been removed. This can be explained in the framework of strain-tuned temporal and spatial fluctuations of f-electron configurations, which might be generally applied to other mixed-valence materials. We note that this amount of strain can be indued in epitaxial SmB$_6$ films via substrate in potential device applications.
\end{abstract}

\maketitle

SmB$_6$ is a prototypical Kondo insulator, a material where the quantum mechanical hybridization between the conduction and $f$-electronic orbitals leads to an opening of a narrow Kondo gap near the Fermi energy \cite{Fisk1996,Riseborough2000}. Experimental observations of the robust surface conductivity on the background of the fully insulating bulk \cite{WolgastFisk2013,KimXia2013} and quantum oscillations from a 2D surface \cite{LiLi2014}, both contribute to the highly nontrivial nature of an insulating state in SmB$_6$. SmB$_6$ has so far attracted much attention as a prominent candidate material for a strongly correlated topological insulator \cite{dzero2010,dzero2012,Alexandrov2013}.  It is noteworthy to mention that the first application of SmB$_6$ as a radio-frequency micro-oscillator device \cite{Stern2016} has been recently demonstrated, operating at cryogenic temperatures.

\begin{figure}
\centering
\includegraphics[width=12cm]{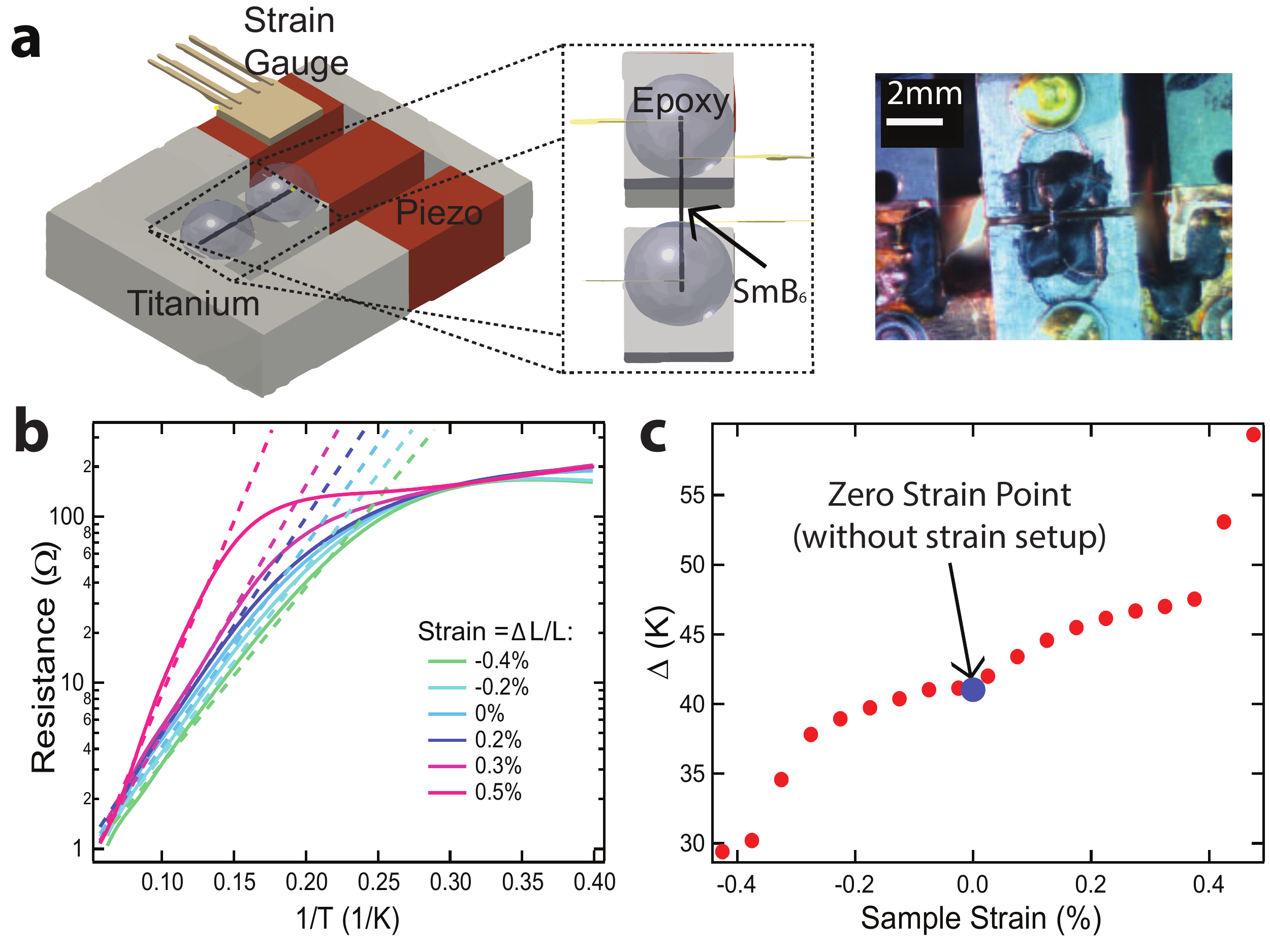}
\caption{\textbf{Experimental setup and gap enhancement with tensile strain in sample A} (a) Sketch and photo of the strain apparatus with SmB$_6$ sample (more details in supplemental material). (b) The resistance of a $\langle$100$\rangle$ oriented crystal as a function of inverse temperature when subject to a relatively small strain between $-0.4\%$ and $0.5\%$. $\Delta$ is measured from the slope of the curve in the bulk dominated conduction region. (c) The extracted bandgap $\Delta$, which increases with tensile strain and decreases with compressive strain. The blue dot is the $\Delta$ of the strain-free state before the sample is mounted to the stain setup. }
\label{Fig1}
\end{figure}

Due to the small value of the band gap $\Delta$, scientific investigations and device applications of SmB$_6$ have thus far been limited to temperatures below the resistance saturation temperature $T\textsuperscript{*} \approx$ 4 $\textit{K}$ when the surface conduction becomes larger than the bulk conduction \cite{WolgastFisk2013,KimXia2013}, motivating the search for means of enhancing $T\textsuperscript{*}$. In the context of thin films, it is known that strain induced by a substrate can dramatically change the temperature scale of the physical phenomena, such as the room temperature ferroelectricity in strained BaTiO$_3$ \cite{Choi2004}. In SmB$_6$, it has been shown that $\Delta$ can be reduced and eventually quenched \cite{Cooley1995} with an external pressure of up to 50 $\textit{kbar}$. Consequently, $T\textsuperscript{*}$ decreases to zero under pressure \cite{Cooley1995, Derr2008}.  A comprehensive theoretical explanation of this intriguing phenomenon has not been reported yet, although it has been speculated to be related to the pressure-dependent valence of Sm ions \cite{Cooley1995}. A related question of whether negative pressure (tensile strain) would alter $\Delta$ or $T\textsuperscript{*}$ in SmB$_6$ has also not been addressed yet.

Applying a large amount of tensile strain in a controllable fashion to macroscopic crystals has  been difficult until very recently when Hicks and Mackenzie introduced a tri-piezo technique \cite{Hicks2015} to enhance the superconducting transition temperature of Sr$_2$RuO$_4$ with strain  which is the fractional change of the sample length \cite{Hicks2014}. For this study we have adopted this technique and constructed a strain setup that can apply between $-0.5\%$ and $1\%$ strain to SmB$_6$ crystals. As shown in Fig. \ref{Fig1}a, the piezo stacks and titanium pieces are arranged so that they compensate for differential thermal contraction \cite{Hicks2015}. Naturally needle-shaped SmB$_6$ crystals obtained directly from growth were mounted to the gap between the titanium pieces using hard cryogenic epoxy. Strain gauges were used to measure the length change of the piezos, which was used in turn to estimate the strain in SmB$_6$ using calibration with direct cryogenics microscopic imaging.  At T = 4 $\textit{K}$, we found that more than 70$\%$ of the strain is transferred from the piezos to the SmB$_6$ needle crystals (see Supplemental Materials for details), which is comparable to what has been seen in Sr$_2$RuO$_4$ \cite{Hicks2014}. With a cubic symmetry and a Poisson's ratio of $\eta=-0.0213$ \cite{Goldstein2012}, all dimensions of SmB$_6$ expand with tensile strain, making it quite analogous to negative hydrostatic pressure.

\begin{figure}
\centering
\includegraphics[width=6cm]{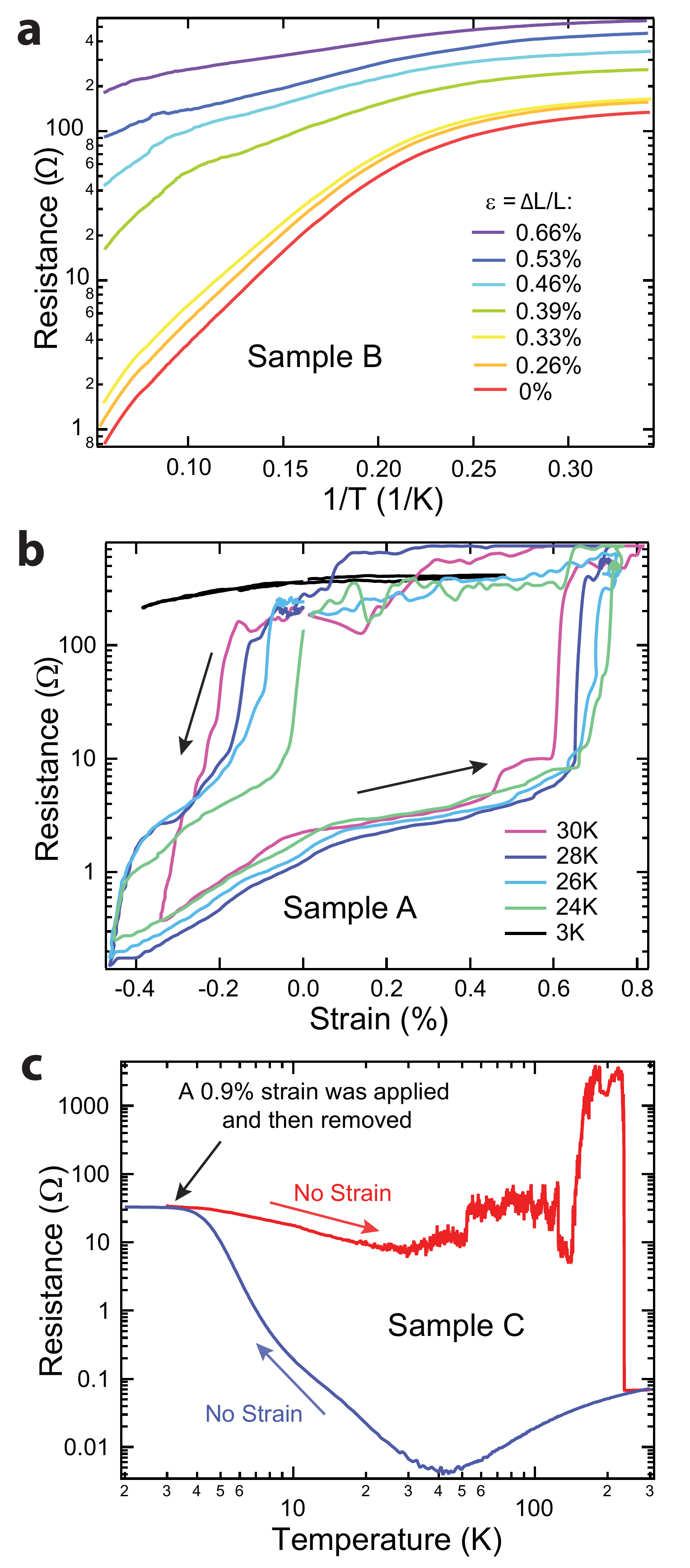}
\caption{\textbf{Hysteresis and room-temperature surface-dominated conduction} (a) The enhancement of $T\textsuperscript{*}$ with tensile strain $\epsilon$ $\leq$ 0.66$\%$ in sample B.  Unlike sample A the low temperature saturation (surface) resistantce in this sample is strain dependent, similar to what was found under pressure in \cite{Cooley1995}. (b) The occurrence of hysteresis with $\epsilon\geq 0.6$$\%$ in sample A. Data shown here represent measurements after the sample has been strained beyond 0.7\%. (c) The persistence of surface dominated conduction up to 240 $\textit{K}$, after a strain of $\epsilon$ =  0.9$\%$ was applied and then removed at low temperature in sample C. } 
\label{Fig2}
\end{figure}

In Kondo insulators the value of an insulating gap is usually obtained by fitting the bulk dominated conduction region to $R^{-1}/R_{\textrm{bulk}}^{-1}= \exp(-\Delta/k_BT)$ \cite{Cooley1995}, where $R_{\textrm{bulk}}$ is a resistance constant for the bulk, $T$ is the temperature, $R$ is the measured resistance, and $\Delta$ is the band gap. As shown in Fig. \ref{Fig1}b for a $\langle$100$\rangle$ oriented SmB$_6$ crystal, when a small compressive strain ($-0.4$\%$\leq\epsilon < 0$) is applied, we found that $\Delta$ is reduced by as much as 30$\%$, comparable to an applied pressure of 24 kbar \cite{Cooley1995}. Given a low-temperature bulk modulus of 910 $\textit{kbar}$ \cite{Tamaki1985} for SmB$_6$, $\epsilon=-0.4\%$ strain is roughly equivalent to 24 $\textit{kbar}$, which again agrees with Cooley et al. \cite{Cooley1995}

In contrast, as we show in Fig. \ref{Fig1}c, a small tensile strain results in exactly the opposite behavior: the resistance saturation temperature $T\textsuperscript{*}$ is shifted to higher temperatures and $\Delta$ is enhanced by 50$\%$ suggesting it has a similar origin to the yet unknown mechanism for pressure-induced $\Delta$ reduction \cite{Cooley1995}. We note that changes in the values of $\Delta$  are much sharper when $|\epsilon|\geq$ 0.4$\%$ suggesting a well-defined collective response for both negative and positive strain.

While compressive strain beyond $\epsilon=-0.5\%$ tends to break needle-shaped samples likely due to shear strain, much larger tensile strains can be safely applied. Shown in Fig. \ref{Fig2}a for another $\langle$100$\rangle$ oriented SmB$_6$ crystal, sample B, tensile strains up to 0.66$\%$ continuously increase the resistance saturation temperature $T\textsuperscript{*}$ up to 30 $\textit{K}$. We note that in this sample in contrast with sample A, the surface dominated saturation resistance is strongly dependent on applied strain reminding us of the similarly diverse behaviors observed in high pressure experiments \cite{Beille1983,Moshchalkov1985,Cooley1995,Derr2008} (see Supplement Materials). Surprisingly, when $\epsilon\geq 0.7\%$ we observe a dramatic onset of hysteretic behavior in the resistivity of SmB$_6$. Note that the possibility of piezo hysteresis can be ruled out by careful experimental analysis (see Supplemental Material on discussion of this issue). As shown in 
Fig. \ref{Fig2}b with sample A, at an elevated temperature of 30 $\textit{K}$ and at tensile strains $\epsilon\geq 0.7\%$ the resistance increases suddenly by more than two orders of magnitude from the bulk dominated resistance of 1 $\Omega$ to the surface dominated resistance of $\sim 100 ~\Omega$ indicative of a first-order phase transition. This dramatic phenomenon is hysteretic in nature: when the strain is reduced to 0\% the sample resistance maintains its surface dominated value until a compressive strain is applied. However, after an application of a large tensile strain, the bulk of SmB$_6$ remains insulating with a significantly increased temperature $T\textsuperscript{*}$ even after the strain is removed. 

To measure the upper bound of $T\textsuperscript{*}$  - the temperature below which the surface conduction is dominant - we prepared sample C. This sample has been strained at low temperature with $\epsilon=0.9\%$, the strain was reduced back to zero and the sample was warmed up to 300 $\textit{K}$. It is shown in Fig. \ref{Fig2}c where the blue curve represents the initial cool down without strain and the red curve shows the warm up after the tensile strain has been applied at low temperature. The noise in the resistance value at temperatures between 40 $\textit{K}$ and 240 $\textit{K}$ is quite repeatable between measurements, suggestive of an underlying fluctuations in the system, which will be discussed below. The sample resistance remains to be more than three orders of magnitude larger than the bulk dominated value (blue curve) normally observed at these temperatures until the temperature reaches $T\textsuperscript{*}$ = 240 $\textit{K}$ when it suddenly reverts back to the bulk resistance value of 0.07 $\Omega$. To our knowledge, this $T\textsuperscript{*}$ is over double the observed value in topological Insulator films of  Bi$_2$Se$_3$ \cite{Ren2010} and Bi$_2$Te$_3$Se \cite{He2012}.

Up to this point we have presented experimental observation of two effects: surface dominated transport in SmB$_6$ at temperatures above $200$ $\textit{K}$ and hysteresis in resistivity under an application of tensile strain. To provide the qualitative understanding of the observed phenomena, we recall that it has been well established experimentally \cite{Mizumaki2009,Butch2016} that SmB$_6$ belongs to a class of intermediate valence systems: 
$f$-orbital electronic configuration of samarium ions fluctuates between the $4f^{5}$ and $4f^6$ states on a typical timescale $\tau_{\textrm{ivc}}$ across a fairly wide temperature range. Experiments which probe the samarium valence on a time-scale longer than $\tau_{\textrm{ivc}}$ will observe an intermediate valence state \cite{Varma1976,PeterReview1981}. Experimentally found values of an intermediate valence configuration vary slightly between Sm$^{2.5+}$ and Sm$^{2.6+}$ suggesting that the energy for the corresponding integer valence configurations are comparable with each other, $E(f^5)\sim E(f^6)$.

With a decrease in temperature, emerging hybridization between the $d$-orbital and $f$-orbital states of samarium opens up a scattering channel $4f^6\leftrightarrow4f^5+5d$ leading to an onset of insulating behavior at low temperatures and also to small changes in the intermediate valence configuration \cite{Mizumaki2009}. Since the ionic volume corresponding to the $4f^6$ configuration exceeds the one for the $4f^5$ configuration, the energy difference between the corresponding valence configurations $\Delta E=E(f^6)-E(f^5)$ will increase with pressure the scattering processes $4f^6\leftrightarrow4f^5+5d$ will be suppressed. As a consequence, one expects that the material recovers its metallic properties in the bulk. On the other hand, an application of a tensile strain acts as a negative pressure and therefore should have an opposite effect of enhancing the insulating behavior and promoting stronger hybridization between the $d$- and $f$-orbitals. Our experimental results for the small tensile strain confirm these expectations. 

\begin{figure}
\centering
\includegraphics[width=16cm]{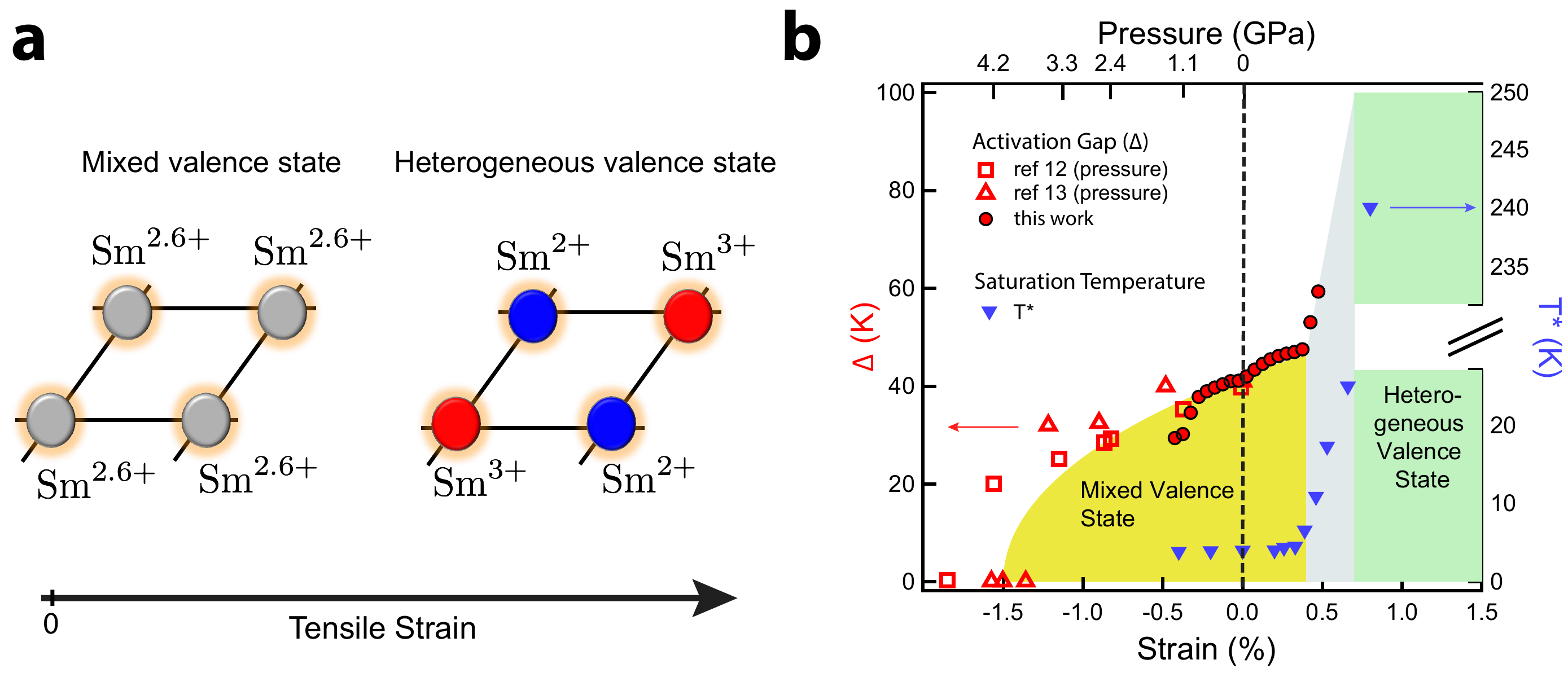}
\caption{\textbf{Temperature-strain phase diagram}  (a) A cartoon illustration of strain-induced transition from temporal fluctuating mixed-valence state to spatially heterogenous valence state. (b) A reconstructed phase diagram of SmB$_6$ under strain with data from this paper as well as pressure data from \cite{Cooley1995, Derr2008}. The left axis corresponds to the Derr bandgap data (hollow red triangles), the Cooley bandgap data (hollow red squares), and our own bandgap data (red circles).  The right axis corresponds to only $T\textsuperscript{*}$, which is shown in blue upside down triangles.  }
\label{Fig3}
\end{figure}

While an increase in $T\textsuperscript{*}$ with tensile strain can be qualitatively understood by the argument based on the changes in the ionic volume, our observation of hysteresis in resistivity as a function of temperature points towards the emergence of the spatial inhomogeneities.  Since the energies of the Sm$^{2+}$ and Sm$^{3+}$ become almost degenerate the characteristic timescale $\tau_{\textrm{ivc}}$ will increase. Ultimately, $\tau_{\textrm{ivc}}\to\infty$ and the system is expected to develop spatially inhomogeneities by becoming mixed-valent: each samarium ion will be in either Sm$^{2+}$ or Sm$^{3+}$ integer valence state, as we have schematically shown in Fig. \ref{Fig3}a. Accordingly we constructed a phase diagram shown in Fig. \ref{Fig3}b: it illustrates the change from mixed to heterogeneous valence state, in comparison with high pressure experiments \cite{Cooley1995, Derr2008}. Note, however, as it was discussed in Ref. \cite{Varma1976}, transition into the spatially inhomogeneous state would be energetically costly since it would mean the departure from the simple rock salt crystal structure. However, in our experiments the energy costs are offset by a tensile strain.  Finally, we note that an abrupt jump in the value of $T\textsuperscript{*}$ signals the first order transition, however its value most likely affected by the local strain and Coulomb interactions between electrons in Sm$^{2+}$ and Sm$^{3+}$ valence states, so that this change in $T\textsuperscript{*}$ may not reflect the corresponding changes in the bulk hybridization gap. 

Most recently, SmB$_6$ has been discussed as a prominent candidate for the first correlated topological insulator. One of the major experimental issues with identifying the topological nature of the metallic surface states has been the smallness of the insulating gap precluding, for example, the precise measurement of the surface electron's chirality. Our experimental findings show that this issue can be, in principle, resolved. On a more general note, our results confirm the basic idea that topological Kondo insulators are likely to be found in systems in the intermediate valence rather than local moment regime

{\textbf{Acknowledgements:} This work was supported by NSF grant DMR-1350122. V.G. acknowledges the support from DOEBES (DESC0001911) and Simons Foundation. M.D. acknowledges the support from NSF (DMR-1506547). We thank S. Thomas, B. Casas and D. Trinh for technical assistance.}

{\textbf{Author Contributions:} A.S. performed the measurements. M.D. and V.G. developed the theory. Z.F. fabricated the samples. J.X. designed the project. All authors discussed the result, and contributed to the writing of the manuscript.}


\newpage

\vskip 7in
\centerline{\Large\bf   
Supplementary Information}

\vskip 0.2in
\centerline{\bf Kondo insulator SmB$_6$ under strain: surface dominated conduction near room temperature}

\newcommand{\beg}{\begin{equation}}
\newcommand{\en}{\end{equation}}
\newcommand{\bp}{\mathbf p}
\newcommand{\bq}{\mathbf q}
\newcommand{\bk}{\mathbf k}
\newcommand{\br}{\mathbf r}
\newcommand{\bR}{\mathbf R}
\newcommand{\bn}{\mathbf n}

\newcommand \bel  {\begin{align}}
\newcommand \enl  {\end{align}}

\newcommand{\eps}{\varepsilon}
\newcommand{\lam}{\lambda}

\newcommand{\re}[1]{(\ref{#1})}
\newcommand{\vecr}{\vec r}
\newcommand{\up}{\uparrow}
\newcommand{\dn}{\downarrow}
\newcommand{\dg}{^\dagger}
\newcommand{\ket}[1]{|#1\rangle}
\newcommand{\bra}[1]{\langle#1|}
\newcommand{\eref}[1]{Eq.~(\ref{#1})}
\newcommand{\rref}[1]{(\ref{#1})}
\newcommand{\esref}[1]{Eqs.~(\ref{#1})}
\newcommand{\pref}[1]{(\ref{#1})}
\renewcommand{\Re}{\mathrm{Re}}
\renewcommand{\Im}{\mathrm{Im}}
\newcommand{\Tr}{\mathrm{Tr}\,}
\bibliographystyle{ieeetr}   

\tableofcontents

\section{Experimental Details}
\subsection{Sample preparation and information}
All samples were grown by the aluminum flux method and HCl was used to remove any aluminum flux or impurities on the samples \cite{Kim2012}.  Samples were selected based on their dimensions, which was possible due to the abundance of SmB$_6$ samples at our disposal.  Long and thin samples were the desired shape so that the strain could be applied homogeneously.  Samples were then polished to remove any surface cracks, which can cause the samples to break prematurely.  Samples were screened prior to being mounted in the strain setup to ensure that samples had the expected resistance to temperature relationship for SmB$_6$ of at least three orders of magnitude change over the temperature range from 300 $\textit{K}$ to 2 $\textit{K}$ \cite{KimXia2013}, which will resolve any uncertainty in locating the zero strain point as we will discuss later.  

Shown below is the table of the dimensions of all samples used in this paper:
\begin{center}
  \begin{tabular}{ | c || r | r | r | r |}
    \hline
    Sample: & Length: & Width: & Height: & $\Delta$V at 0.5\% Strain: \\ \hline
    A & 2537$\mu$m & 287$\mu$m & 161$\mu$m & 0.5364\% \\ \hline 
    B & 4448$\mu$m & 456$\mu$m & 208$\mu$m & 0.5342\% \\ \hline 
    C & 4032$\mu$m & 303$\mu$m & 156$\mu$m & 0.5315\% \\ \hline 
  \end{tabular}
\end{center}
The dimensions are all accurate to within 10$\%$.  We found that samples that were less than 2mm long had difficulty being held by the epoxy. Sample B had the largest cross-sectional area and was also the hardest sample to apply strain to, while sample A was the easiest sample to strain due to its small cross-sectional area.  Smaller cross-sectional area should result in more uniform strain \cite{Hicks2014}.  The change in volume caused by a strain is roughly 6\% greater than the strain applied, due to a Poisson's ratio of -0.0213 \cite{Goldstein2012} when strain is applied in the $\langle$100$\rangle$ direction.

We chose to mount the samples with epoxy over a mechanical connection/clamping mechanism because mechanical connection/clamping would add additional pressure to the crystal, which would increase the risk of breaking or cracking. Applying an additional pressure to the crystals would also make to strain not purely uniaxial and potentially affect the experimental results. Several other experiment used epoxy successfully to apply strain, including \cite{Koduvayur2011}, \cite{Chu2012}, and \cite{Hicks2014}. Samples were mounted using Stycast 2850FT epoxy and catalyst 24 LV.  Then, the epoxy was cured at room temperature for at least 24 hours, which reduced the risk of thermal expansion affecting the experiment.  This epoxy is harder than most epoxies \cite{Hicks2014}, allowing for strain to be more efficiently applied.  

\subsection{Transport measurements}

We measured resistance using a Signal Recovery 7225 DSP lock-in amplifier. The corresponding gap $\Delta$ is obtained by fitting the bulk dominated conduction region to $R = R_{\textrm{bulk}} \exp(\Delta/k_BT)$, where, $R_{\textrm{bulk}}$ is a resistance constant for the bulk, T is the temperature, R is the measured resistance, and $\Delta$ is the band gap \cite{Cooley1995}.  The bulk dominated conduction temperature range where the fitting is done is between 4 $\textit{K}$ and 20 $\textit{K}$, when the strain is small.  At higher strains, the fitting temperature range needs to be adjusted due to the changing $T\textsuperscript{*}$.

Four platinum wires were attached to the sample using spot-welding or silver epoxy, with no measurable difference caused by the different methods.  Spot-welding uses a large current over a small area and a short period of time to melt a small wire on to the sample.  This technique works effectively on metallic samples producing a very low contact resistance \cite{Kim2012}.

\begin{figure}
\centering
\includegraphics[width=10cm]{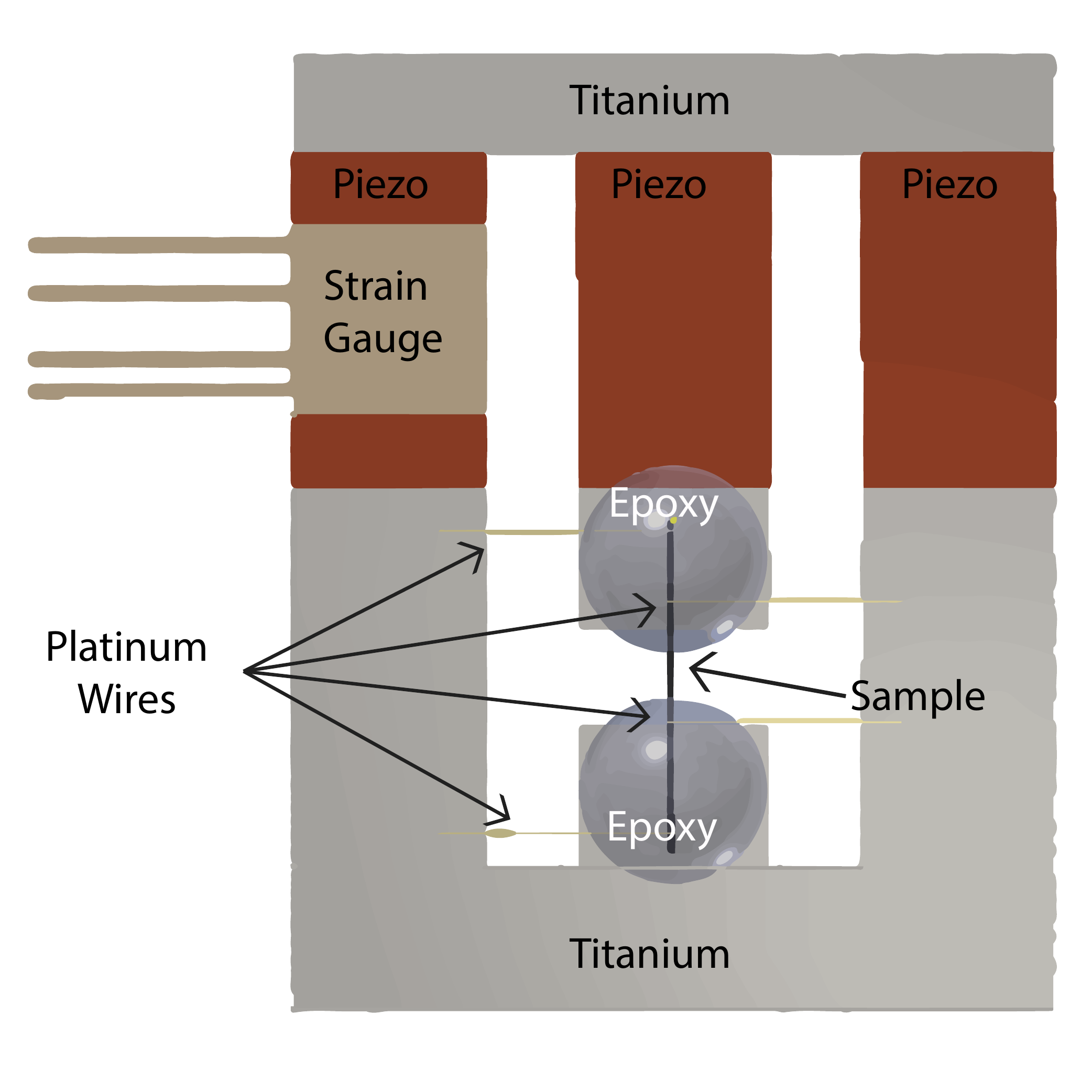}
\caption{\textbf{Strain setup design.} The design is meant to minimize the effects of differential thermal contraction. The sample is held in place by Stycast 2850FT epoxy.}
\label{Schematic}
\end{figure}

The strain setup design is shown in Fig. \ref{Schematic}, following the design of Hicks et. al \cite{Hicks2015}, which limits the amount of residual strain applied to the sample from differential thermal contraction.  The thermal contraction of the piezos is quite large, but their arrangement should, in theory, cancel each other.  The body of the device is made of titanium, which has a very low thermal contraction. From our calculations, this would give us no net strain from cooling.  This design differs from the Hick and Mackenzie setup because it uses larger piezos allowing us to apply more strain.

All strain measurements were done in a Janis SHI-4-2 Closed Cycle Optical Cryostat.  A large thermal mass was added to the cold stage to help stabilize the temperature to less than 0.05 $\textit{K}$ fluctuations, with a minimum temperature of 2.5 $\textit{K}$.

\begin{figure}
\centering
\includegraphics[width=14cm]{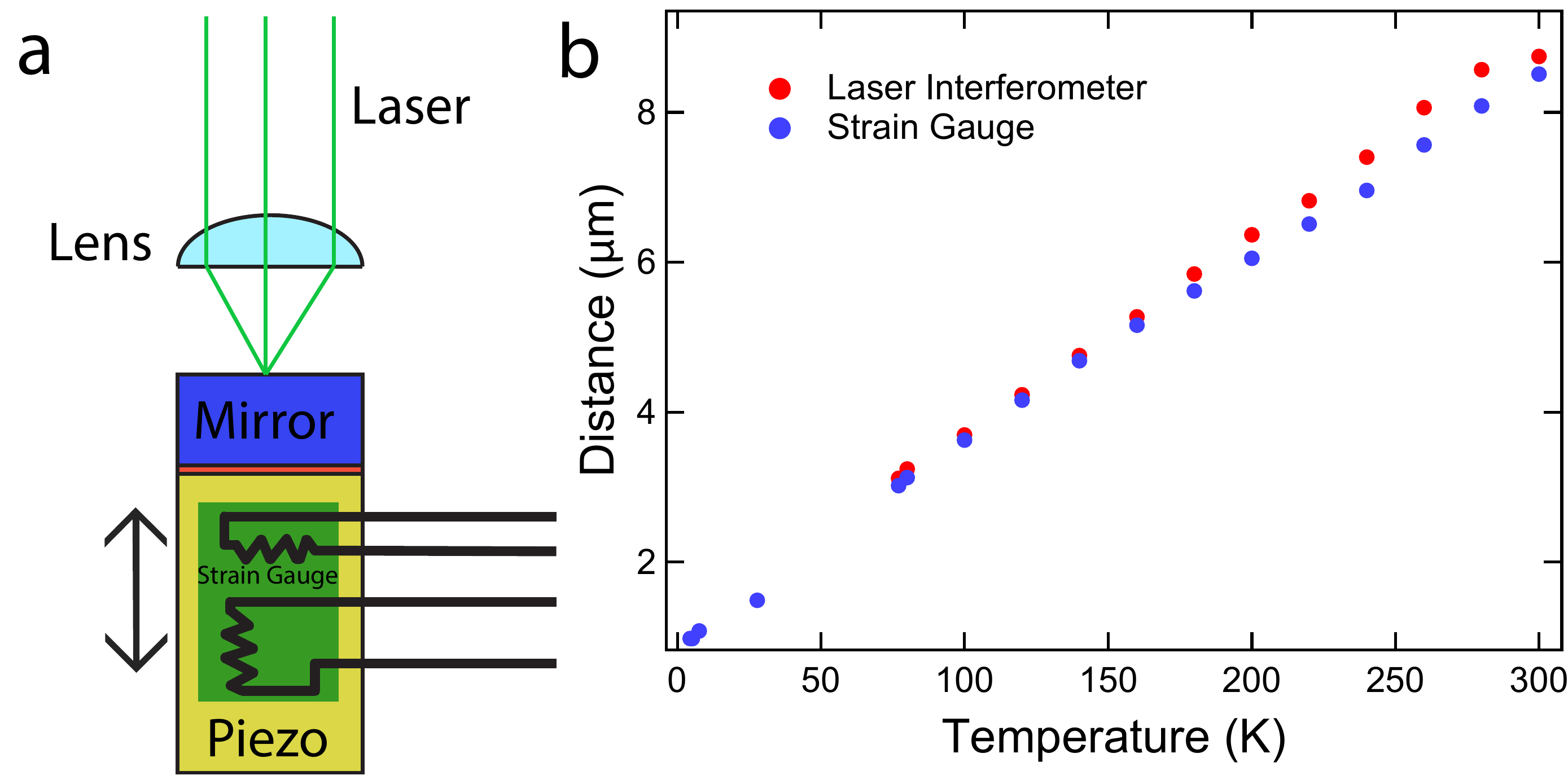}
\caption{\textbf{Laser Interferometer to strain gauge calibration.} {\bf a,} The laser interferometer is used to measure the total displacement of the piezo by reflecting the laser off of a mirror on the top of the piezo.  A strain gauge is mounted to the side of the piezo using Stycast 2850FT epoxy to measure the piezo strain.   {\bf b,} These two measurements are converted to a total distance movement to be compared between liquid helium temperature and room temperature.}
\label{Fig2}
\end{figure}

\subsection{Strain calibration}
To verify the strain is actually being applied with our setup, we used a laser interferometer to calibrate the strain gauges on the piezos and then used image comparison software of the sample to compare the strain gauge to the strain applied.  Our goal was to use the readings from the strain gauges mounted on the piezos to verify the actual strain applied to the sample.

We calibrated the strain gauges using a laser interferometer.  Two strain gauges were mounted on the same side of a piezo using Stycast 2850FT epoxy, with one strain gauge set to measure any movement and the other set to measure any movement perpendicular to the primary axis of movement, which should be minimal. The strain gauges are arranged into a resistance bridge to eliminate any non-strain fluctuations, such as temperature.  The strain gauges were measured with a lock-in amplifier so that we could use a smaller current of 10 $\mu$A.  We noticed that the strain gauges could add a significant heat load to the strain setup when using currents close to 10mA, which is partly due to the low thermal conduction of the piezos.  We then used the interferometer to make sure that the movement being measured by the strain gauges accurately matched the movement of the piezo by mounting a mirror to the top of the piezo, shown in Fig. \ref{Fig2}(a).  The data are shown in Fig. \ref{Fig2}(b) with at least 94$\%$ of the laser interferometer movement measured by the strain gauges. The data also verified the manufacture specifications for the piezos by a laser interferometer and a strain gauge.

\begin{figure}
\centering
\includegraphics[width=14cm]{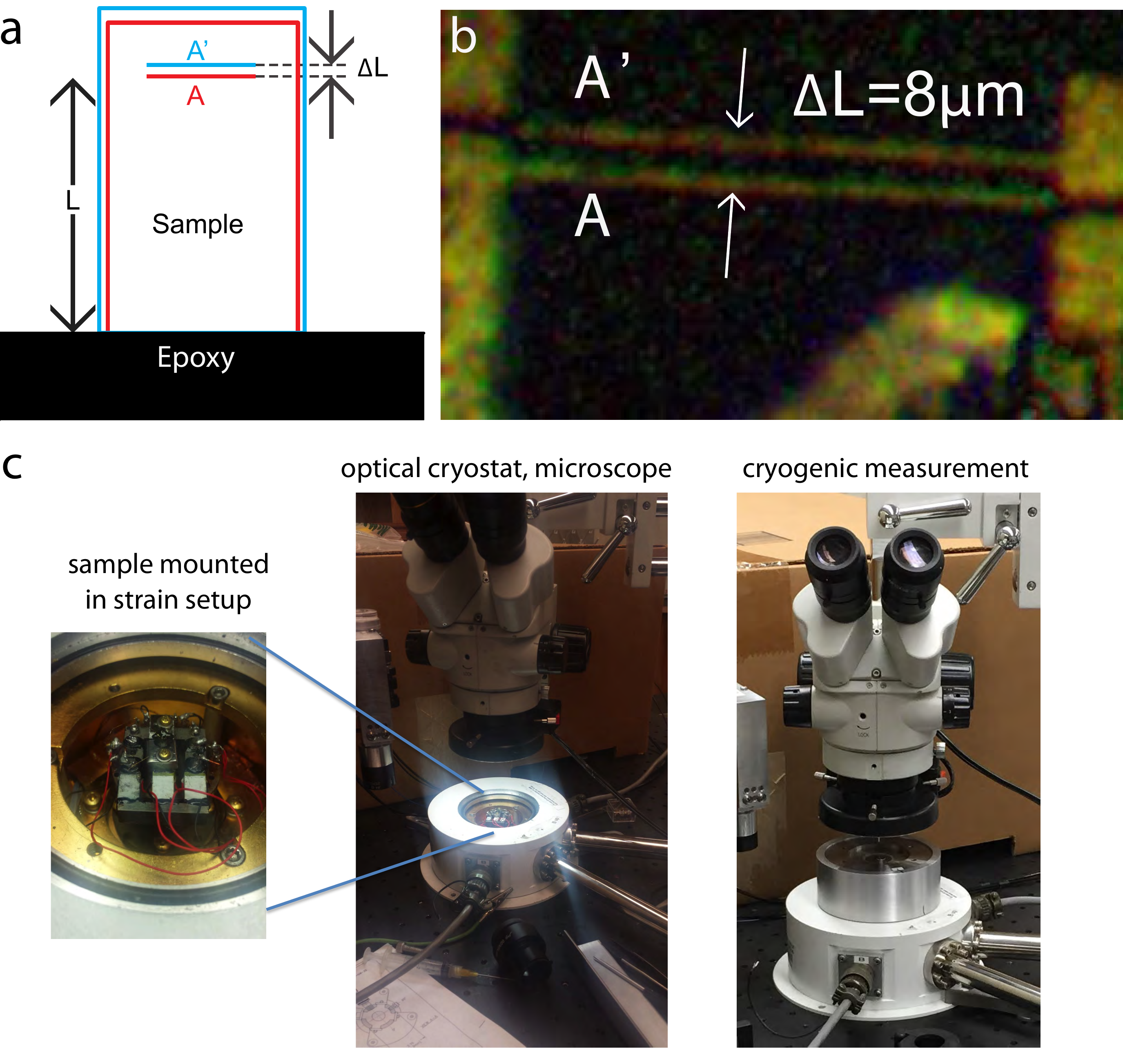}
\caption{\textbf{Direct measurement of sample strain using a microscope at cryogenic temperatures.} {\bf a,} The cartoon representation for measure strain using image comparion.  The bottom end does not move and the top end is connected to the movable end of the strain setup.  Both ends are mounted in epoxy, but the top end is not shown.  We now compare the movement of a marker on the sample. {\bf b,} This image shows the actual comparison of two separate images: one strained and the other unstrained at liquid helium temperature.  The change in the placement of the line can be used to calculate the the stress applied to the sample. {\bf c,} The strain set-up is mounted in the cryostat with a small window above the sample.  A microscope with a camera is placed above the window to take photos.}
\label{Fig3}
\end{figure}

While the laser interferometer verified the piezo movement, this can only be used to verify the displacement of piezo, not the sample strain.  The displacement of piezo includes the strain absorbed by the epoxy and the sample, while the sample strain is proportional to the change in length of just the sample. 

Hicks et. al was able to show that Stycast 2850FT is able to be used to apply strain to a sample of S$_2$RO$_4$ \cite{Hicks2014}. The pioneering work Hicks et. al used numerical simulations to convert the total displacement to sample strain showing that 73\% of the strain is transferred to the sample, with the rest absorbed by the expoxy. SmB$_6$ has a smaller bulk modulus than S$_2$RO$_4$ \cite{Tamaki1985}, so it will be easier to apply strain. The numerical simulation might not have been as accurate for larger strain. Therefore, we developed a new technique by taking images of our samples and comparing features to find the sample strain.  We were able to do this because we were using higher strain.

Shown in Fig. \ref{Fig3}(c) is the optical strain set-up, which works between room temperature and liquid helium temperature.  A microscope with a camera is mounted outside of the optical cryostat and above the sample to take images of the sample under strain.  One images are taken of the strained sample and one when the sample is not strained.  Fig. \ref{Fig3}(a) shows a cartoon representation of the sample stretching.  We pick a feature on the sample and then track its position change.  We compared an unstrained sample to itself being strained shown in Fig. \ref{Fig3}(b).  The lower line is the original feature, while the top line is the same feature on the same sample, but with the sample being stretched.  The image shows a 9 pixel change or 0.8\% applied stress at liquid helium temperature.  Using all of this data, we can show that approximately 70\% of the strain is transferred to the sample, which is comparable to other similar experiments \cite{Hicks2014}.  All of the values of strain in the paper were measured by the strain gauge, not determined by the voltage applied to the piezos.

\subsection{Negligible residual strain}

\begin{figure}
\centering
\includegraphics[width=10cm]{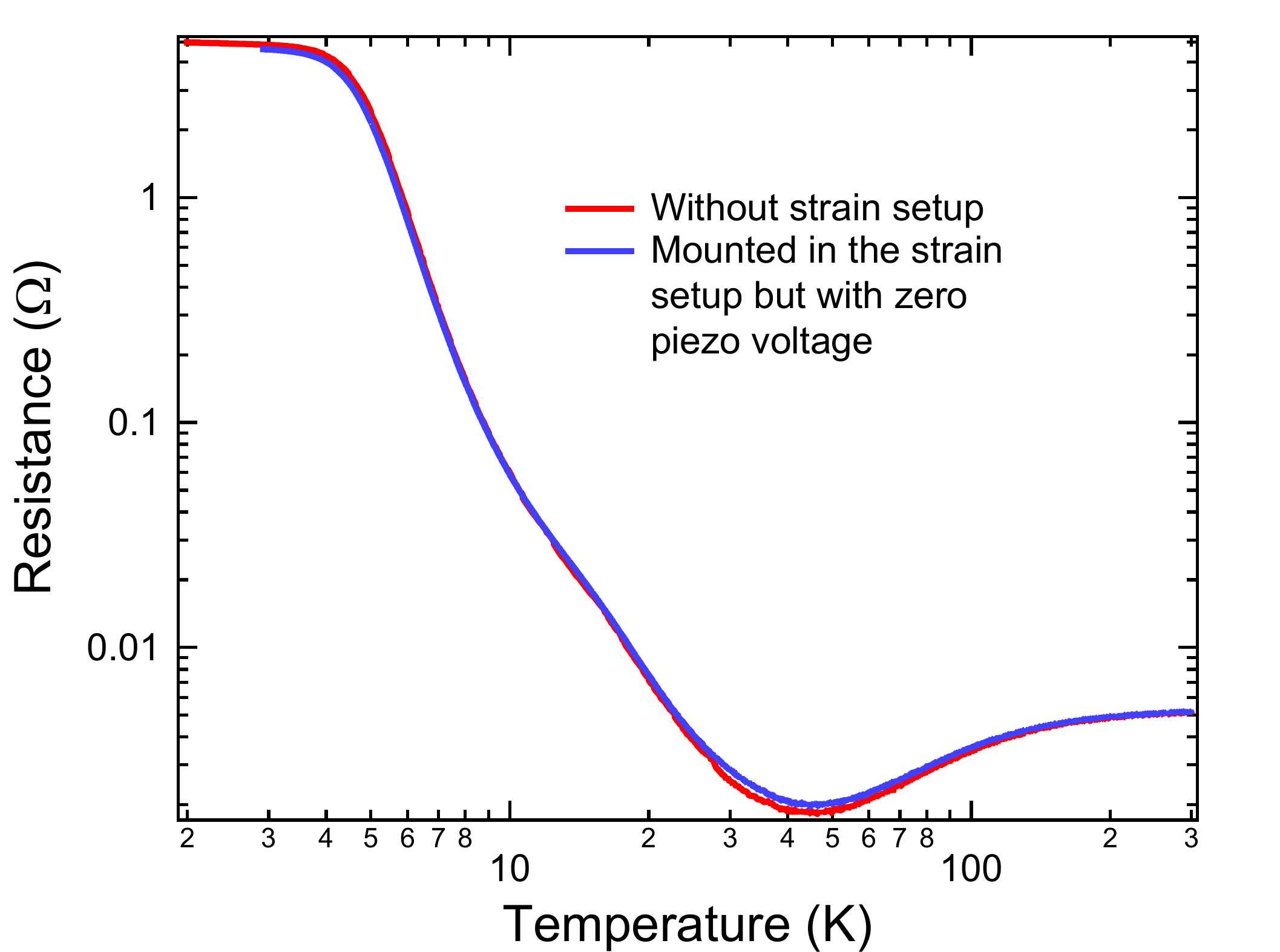}
\caption{\textbf{Setup Strain applied.} The measured resistance of a sample before and after being mounted in our strain setup.} 
\label{Fig4}
\end{figure}

To verify that the samples were not actually being strained during the cooling process, we tested some of the samples' resistance versus temperature relationship before being mounted in the strain setup and after being mounted in the strain setup.  The unmounted sample was measured in a Physical Properites Measurement System (PPMS) by Quantum Design with the sample dangling by the wires so that no strain is applied to the sample.  We then measured the sample's resistance versus temperature with no voltage applied to the piezos.  We then verified that the resistance results matched, shown in Fig. \ref{Fig4}.  The results match within the margin of error of the measurement tool, showing only negligible strain applied from 300 $\textit{K}$ to 2.5 $\textit{K}$ with no voltage applied across the piezos.

\section{Comparison to prior experiments}
\subsection{Conversion}

To compare our strain result to prior high pressure experiments, we used an approximate conversion based on volumetric change \cite{Cooley1995, Derr2008}.  We compared our results to Cooley and Derr by converting their results into a volumetric change and then using our specific sample's volume change to convert the Cooley and Derr values from volumetric change to strain.  This should give us a general interpretation of our results.

The differential form for the change in volume shown in (\ref{diff V}), with a more explicit form shown in (\ref{volume change}).  
\beg \label{diff V}
\frac{dV}{V} = \frac{-dP}{K}
\en

\beg \label{volume change}
V = V_0 \exp(-P/K)
\en

Here, V is the volume, K is the bulk modulus, and P is the pressure.  SmB$_6$ has a bulk modulus of 910 $\textit{kbar}$ \cite{Tamaki1985} at low temperatures for SmB$_6$.  We calculated the volume changes for each of the Derr and Cooley results.

We then calculated the volumes change as a result of strain using Poisson's ratio.  Our samples are rectangular bars with a volume from (\ref{Volume}), where L is length, W is width, and H is height.
\beg \label{Volume}
V = L*W*H
\en

By applying a small strain, or $\Delta$L, the crystal dimensions all change slightly, which changes the volume.  SmB$_6$ has a negative Poisson's ratio of -0.0213 \cite{Goldstein2012}, meaning that the cross-sectional area increases with strain and the volume increase is more than just the result of the strain applied. The formulas for the new dimenions and volume are shown in (\ref{Poisson ratio Volume change},\ref{Width Change},\ref{Height Change}), where $\nu$ is the Poisson's ratio.

\beg \label{Poisson ratio Volume change}
V + \Delta V = (L+\Delta L) (W+\Delta W) (H+\Delta H)
\en

\beg \label{Width Change}
\Delta W = \nu W \frac{\Delta L}{L}
\en

\beg \label{Height Change}
\Delta H = \nu H \frac{\Delta L}{L}
\en

We calculated the volume changes for each comparable sample and instead of using it to convert the strain into a volume change, we used it as a conversion factor between the pressure induced volume change and strain.  This is shown in Fig. 3(b) in the main text.

\subsection{Previous high pressure results}

There have been four previous studies on SmB$_6$ under high pressure, which we examined. All of the studies showed that the insulating behavior collaspes into a metal with sufficiently high pressure, but there are several key differences.  Beille found that the gap closed around 62 $\textit{kbar}$ and that the surface resistance is affected by pressure \cite{Beille1983}.  Moshchalkov found a similar result, with the gap closing around 57 $\textit{kbar}$ \cite{Moshchalkov1985}.  Then, Cooley found the gap closing around 53 $\textit{kbar}$, which was attributed to higher sample quality.  More recently, Derr repeated this experiment using a Bridgman anvil cell and a diamond anvil cell, both of which showed minimal surface resistance change with pressure.  The Bridgman anvil cell showed the gap closing around 39 $\textit{kbar}$ and the diamond anvil cell showed the gap closing aroung 105 $\textit{kbar}$ \cite{Derr2008}.  Derr proposed that the discrepancy was from the uniformity of the pressure application with the diamond anvil cell superior in uniformity. Our samples were grown is separate batches, but measured with the same strain device, suggesting that the biggest reason for variation is the sample quality, in agreement with Cooley.

\section{Theory details: fluctuations in Kondo insulators under a tensile strain}
In the main text we have demonstrated that the application of a tensile strain to samarium hexaboride gives rise to two main effects: (i) the substantial increase in the coherence temperature $T^*$, which is equivalent to an increase of the band gap and (ii) hysteretic behavior in transport. We have argued that a gradual increase in tensile strain enhances the valence fluctuations and, in addition, brings the system close to the first order transition from an intermediate valence state into a heterogeneous valence state with broken translational symmetry. In particular, 
it appears that the application of a tensile strain leads to an increase in an intervalence configuration time $\tau_{\textrm{ivc}}$. 
In this Section of the Supplementary Materials we will address the physics related to a sharp increase in $T^*$ and the role of the valence fluctuations under strain.  Specifically, we consider the effect of fluctuations on the $f$-electron self-energy $\Sigma_{ff}$ as well as on the $f$-electron occupation number $n_f$ within the mean-field slave-boson approximation for the Anderson lattice model. We find that the fluctuation correction to $n_f$ decreases with an increase in tensile strain. Furthermore, when the magnitude of the tensile strain exceeds a certain value, the contribution of the fluctuations to the $f$-electron relaxation time determined by the imaginary part of an $f$-electron self-energy, $\hbar/\tau_{ff}= \textrm{Im}\Sigma_{ff}$, decreases as the value of strain increases. 

\subsection{Model} 
Since we are primarily interested in finding the fluctuation correction to the $f$-electron number, we will ignore the dispersion for the $f$-electrons and restrict the analysis to the Anderson lattice model in two-dimensions.  In what follows we will adopt the functional integral formalism in which the partition function has the usual form:
\beg\label{PFunction}
{\cal Z}=\int{\cal D}[\overline{c},c,\overline{f},f]e^{-S}, \quad S=\int\limits_0^\beta d\tau L(\tau), 
\en
and the Lagrangian $L(\tau)$ is given by:
\beg\label{SFull}
\begin{split}
L(\tau)&=
\sum\limits_{\bk \sigma}\left\{\overline{c}_{\bk \sigma}(\tau)(\partial_\tau+\varepsilon_\bk)c_{\bk \sigma}(\tau)+
\overline{f}_{\bk \sigma}(\tau)(\partial_\tau+\epsilon_f)f_{\bk \sigma}(\tau)\right\}\\&+V\sum\limits_{j\sigma}\left[\overline{c}_{j \sigma}(\tau)f_{j\sigma}(\tau)+\overline{f}_{j\sigma}(\tau)c_{j\sigma}(\tau)\right]+U_{ff}\sum\limits_i 
\overline{f}_{j\up}(\tau){f}_{j\up}(\tau)\overline{f}_{j\dn}(\tau){f}_{j\dn}(\tau),
\end{split}
\en
where $\beta=1/k_BT$, $\overline{c},c$ are fermionic fields which describe the electrons in the conduction band with dispersion $\epsilon_\bk$, 
$\overline{f},f$ are the fermionic fields which account for the $f$-electrons with single particle energy $\epsilon_f$,  
$\sigma=(\up,\dn)$,  ${V}$ is the hybridization amplitude and $U_{ff}$ is the repulsive coupling between the $f$-electrons. 

\begin{figure}[h]
\centering
\includegraphics[width=9cm,height=12cm,angle=-90]{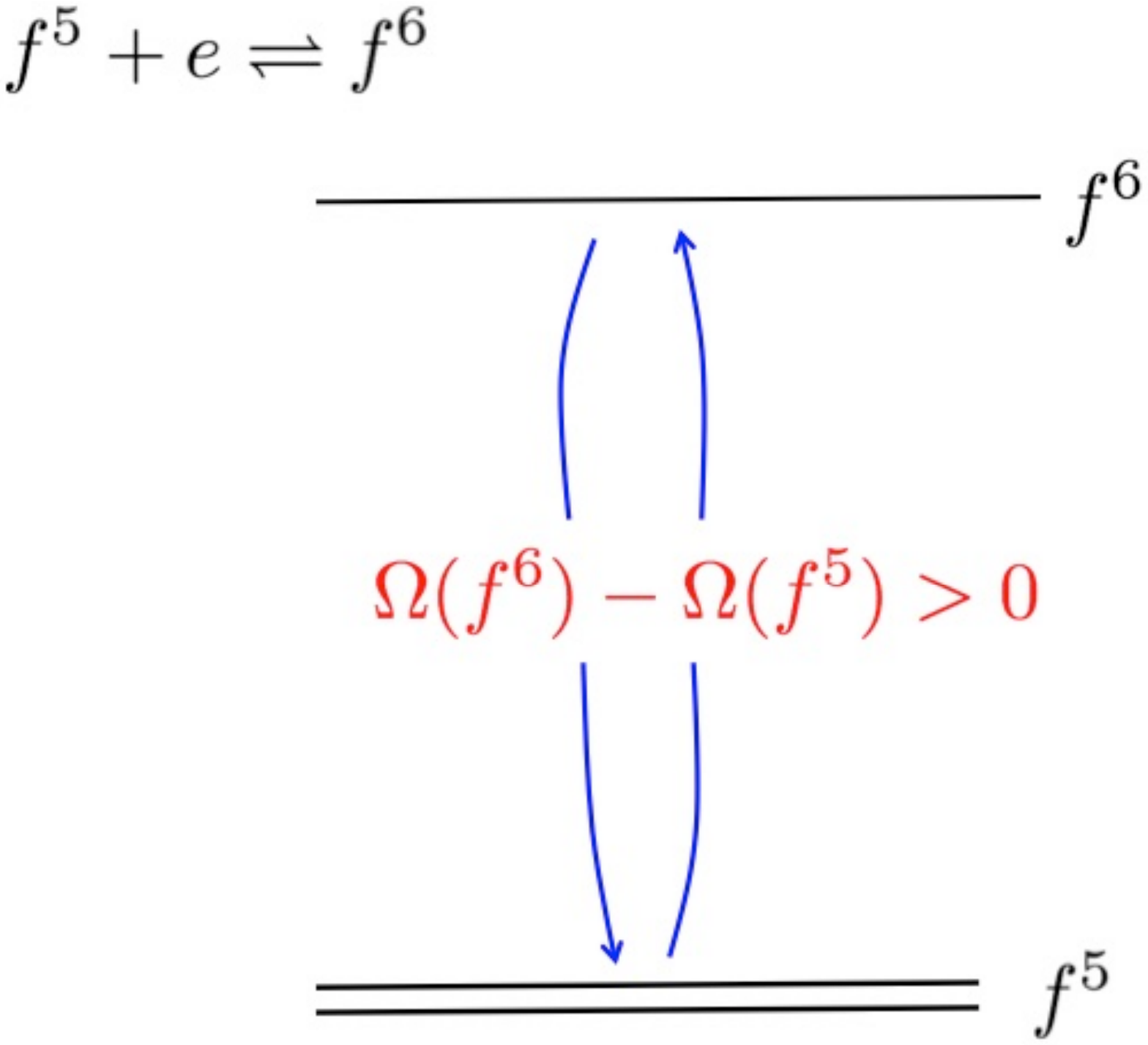}
\caption{\small(Color online) Schematic presentation for the virtual charge fluctuations between magnetic $f^5$ and non-magnetic $f^6$ valence configurations:  $f^6\rightleftharpoons f^5+e$. Since the ionic volume of the $f^6$ configuration is larger than the ionic volume of the $f^5$
configuration, an application of a tensile strain enhances the valence transitions between these two configurations.}
\label{FigVolume}
\end{figure}

\paragraph{Effects of strain.} In what follows we assume that the main effect of tensile or compressive strain is limited to the changes in the local volume of the $f$-ions.  Let us set $\Omega_t$ to be the total ionic volume of, say, samarium ion. We will also ignore the effects of anisotropy which are introduced by the application of stain. Then, at high temperatures when there is no coherence between conduction and $f$-electrons, the ionic volume must be equal to the one corresponding to the magnetic valence configuration, $\Omega_t=\Omega(f^5)$, since the ground state valence configuration for samarium ion is Sm$^{2+}$.
As temperature is decreasing, hybridization between the conduction and $f$-electrons leads to an opening of the hybridization gap governed
by the quantum mechanical transitions between the $f^5$ and $f^6$ valence configurations, Fig. \ref{FigVolume}. 
The opening of the hybridization gap can be described by the emergence of the non-zero expectation value 
$v=\sum\limits_\sigma\langle\overline{c}_{j\sigma}f_{j\sigma}\rangle$ at each lattice site. Then, for the ionic volume one obtains \cite{Zhang2002}:
\beg\label{tVolume}
\Omega_t=\Omega(f^5)(1-|v|^2)+\Omega(f^6)|v|^2=\Omega(f^5)+\delta\Omega |v|^2.
\en
with $\delta\Omega=\Omega(f^6)-\Omega(f^5)$.

\subsection{Mean-field approximation} The hybridization process can be effectively described within the slave-boson mean-field approximation for the Anderson model (see e.g. \cite{NewnsRead1987}).  In the slave-boson approximation one first takes the limit $U_{ff}\to\infty$, which corresponds to projecting out the doubly occupied states and then one introduces the constraint which guarantees the local moment at the $f$-site, i.e. $n_f=1$: 
\beg\label{constraint}
\begin{split}
& U_{ff}\to\infty: \quad f_{i\alpha}\to f_{i\alpha}b_i\dg, \quad f_{i\alpha}\dg\to f_{i\alpha}\dg b_i, \quad
\sum\limits_{\alpha}f_{i\alpha}\dg f_{i\alpha}+
b_i\dg b_i=1,
\end{split}
\en
where $b_i$ are bosonic fields which project out the doubly occupied $f$-states. 
Therefore, within the slave-boson approximation the Lagrangian which accounts for the effect of pressure or tensile strain reads
\beg\label{LV}
L_{\Omega}(\tau)=-P\delta\Omega \sum\limits_{j}\overline{b}_j(\tau)b_j(\tau),
\en
where $P$ is an external pressure. 
Thus, the application of pressure leads to a shift of the $f$-level energies. The resulting effective action for the problem can be formulated in terms of the slave fields only. It follows
\beg\label{Seff}
\begin{split}
Z&=\int{\cal D}[\rho,\lambda]e^{-S_{\textrm{eff}}}, \\
S_{\textrm{eff}}&=-\textrm{Tr}\ln\hat{A}+iT^2\sum\limits_{kk'}\rho(-k')\lambda(k'-k)\rho(k)-\frac{i}{2}\int\limits_{0}^\beta d\tau\sum\limits_j\lambda_j(\tau)\\&+\int\limits_0^\beta d\tau\sum\limits_{j}\rho_j(\tau)\partial_\tau\rho_j(\tau)-PT\delta\Omega \sum\limits_{k}
\rho(-k)\rho(k).
\end{split}
\en
where $k=(\bk,i\omega)$, $\rho_j=|b_j|$, $\lambda_j$ is a constraint field and matrix $\hat{A}$ is defined according to 
\beg\label{MatrixA}
A_{kk'}=\left[\begin{matrix}
(-i\omega+\varepsilon_\bk)\delta_{kk'} & TV\rho(k-k') \\ TV\rho^*(k-k') & (-i\omega_n+\epsilon_f)\delta_{kk'}+iT\lambda(k-k')
\end{matrix}\right].
\en

In the mean-field approximation, all slave-boson operators as well as constraint fields at each site are replaced with $c$-numbers:
\beg\label{mf}
\rho_\bk(\tau)=\sum\limits_j\rho_j(\tau)e^{-i\bk\cdot\br_j}=\overline{\rho}\delta_{\bk,0}, \quad 
i\lambda_\bk(\tau)=(E_f-\epsilon_{f})\delta_{\bk,0}=i\overline{\lambda}\delta_{\bk,0},
\en
where $\epsilon_{0f}$ is the bare $f$-level position. The action at the mean-field level becomes
\beg\label{Smf}
\begin{split}
S_0&=-\textrm{Tr}\ln\hat{A}_0+\frac{1}{T}\overline{\rho}^2\left(i\overline{\lambda}-P\delta\Omega\right)-\frac{1}{T}i\overline{\lambda},\\
[\hat{A}_0]_{kk'}&=\left[\begin{matrix}
(-i\omega+\varepsilon_\bk)\delta_{kk'} & V\overline{\rho}\delta_{kk'} \\ V\overline{\rho}\delta_{kk'} & (-i\omega_n+\epsilon_f)\delta_{kk'}+i\overline{\lambda}\delta_{kk'}
\end{matrix}\right]\equiv-[\hat{G}_0^{-1}]_{kk'}
\end{split}
\en
so that the bare Green's function can formally written as
\beg\label{G0}
\hat{G}_0(k)=\left[\begin{matrix}
G_0^c(k) & G_0^m(k) \\ G_0^m(k) & G_0^f(k)
\end{matrix}\right].
\en

Mean-field equations are found by find an extremum of the action $S_0$ with respect to $(\overline{\rho},i\overline{\lambda})$ and the condition for the particle number conservation. It is straightforward to check that the application of the tensile strain corresponding to $P\delta\Omega>0$ necessarily leads to a linear in $P\delta\Omega$ increase of $\overline{\rho}$, which determines the hybridization gap \cite{Zhang2002,Dzero2014SCES}. 

\subsection{Fluctuations}
Consider the fluctuation correction to the mean-field:
\beg
\rho(k)=\frac{\overline{\rho}}{T}\delta_{k,0}+\tilde{\rho}(k), \quad 
i\lambda(k)=\frac{i\overline{\lambda}}{T}\delta_{k,0}+i\tilde{\lambda}(k).
\en
For the effective action we write $S_{\textrm{eff}}=S_0+{S}_{\textrm{eff}}^{(2)}$. There are two contributions to ${S}_{\textrm{eff}}^{(2)}$. The first one originates from the first term in (\ref{Seff}):
\beg\label{As}
\begin{split}
-\textrm{Tr}\ln\hat{A}&=-\textrm{Tr}\ln\hat{A}_0-\textrm{Tr}\ln[1-\hat{G}_0\hat{A}_{\textrm{fl}}], \\
[\hat{A}_{\textrm{fl}}]_{kk'}&=\left[\begin{matrix}
0 & VT\tilde{\rho}(k-k') \\ VT\tilde{\rho}^*(k-k') & iT\tilde{\lambda}(k'-k)\end{matrix}\right].
\end{split}
\en
For the gaussian fluctuations, we expand the expression under the log up to the second order in powers of $\hat{A}_{\textrm{fl}}$.
The linear order term vanishes, while the second order term in the expansion
\beg\label{Gaussian}
-\textrm{Tr}\ln[1-\hat{G}_0\hat{A}_{\textrm{fl}}]\approx
T\sum\limits_q[\tilde{\rho}(-q),i\tilde{\lambda}(-q)]\left[
\begin{matrix}
\Gamma_\rho(q) & \Gamma_m(q) \\ {\Gamma}_m(-q) & \Gamma_\lambda(q)
\end{matrix}\right]
\left[\begin{matrix} \tilde{\rho}(q) \\ i\tilde{\lambda}(q)\end{matrix}\right], 
\en
where
\beg\label{Gammas}
\begin{split}
\Gamma_\rho(q)&=TV^2\sum\limits_{i\omega}\sum\limits_\bk\left[G_0^m(k)G_0^m(k+q)+G_0^c(k)G_0^f(k+q)\right], \\
\Gamma_m(q)&=TV\sum\limits_{i\omega}\sum\limits_\bk G_0^m(k)G_0^f(k+q), \quad \Gamma_\lambda(q)=T\sum\limits_{i\omega}\sum\limits_\bk G_0^f(k)G_0^f(k+q).
\end{split}
\en
Thus, for the fluctuation correction to the effective action at the gaussian level we find
\beg\label{Seff2}
\begin{split}
S_{\textrm{eff}}^{(2)}=T\sum\limits_{\bk}\sum\limits_{i\nu_l}[\tilde{\rho}(-k),i\tilde{\lambda}(-k)]\left[
\begin{matrix}
-i\nu_l-\eta+i\overline{\lambda}+\Gamma_\rho(k) & \overline{\rho}+\Gamma_m(k) \\ \overline{\rho}+{\Gamma}_m(-k) & \Gamma_\lambda(k)
\end{matrix}\right]
\left[\begin{matrix} \tilde{\rho}(k) \\ i\tilde{\lambda}(k)\end{matrix}\right] 
\end{split}
\en
and $i\nu_l=2\pi Tl$ are bosonic Matsubara frequencies. 
\paragraph{Fluctuation propagator.}
Ultimately, we would like to compute the correction to the $f$-electron occupation numbers due to fluctuations. Since only $i\lambda(k)$ couples directly to the $f$-electron particle number, we will need to evaluate the fluctuation propagator
\beg\label{Dlam}
D_\lambda(k,k')=-\langle\hat{T}_\tau i\tilde{\lambda}(\bk,\tau)i\tilde{\lambda}(-\bk',\tau')\rangle.
\en
From (\ref{Seff2}) it immediately follows that
\beg\label{Dlam2}
D_\lambda(\bk,i\nu)=-\frac{i\nu+\eta+\epsilon_f-E_f-\Gamma_\rho(k)}{\Gamma_\lambda(k)[i\nu+\eta+\epsilon_f-E_f-\Gamma_\rho(k)]+[\overline{\rho}+\Gamma_m(k)][\overline{\rho}+\Gamma_m(-k)]}.
\en
\begin{figure}[h]
\centering
\includegraphics[width=12cm,height=9cm]{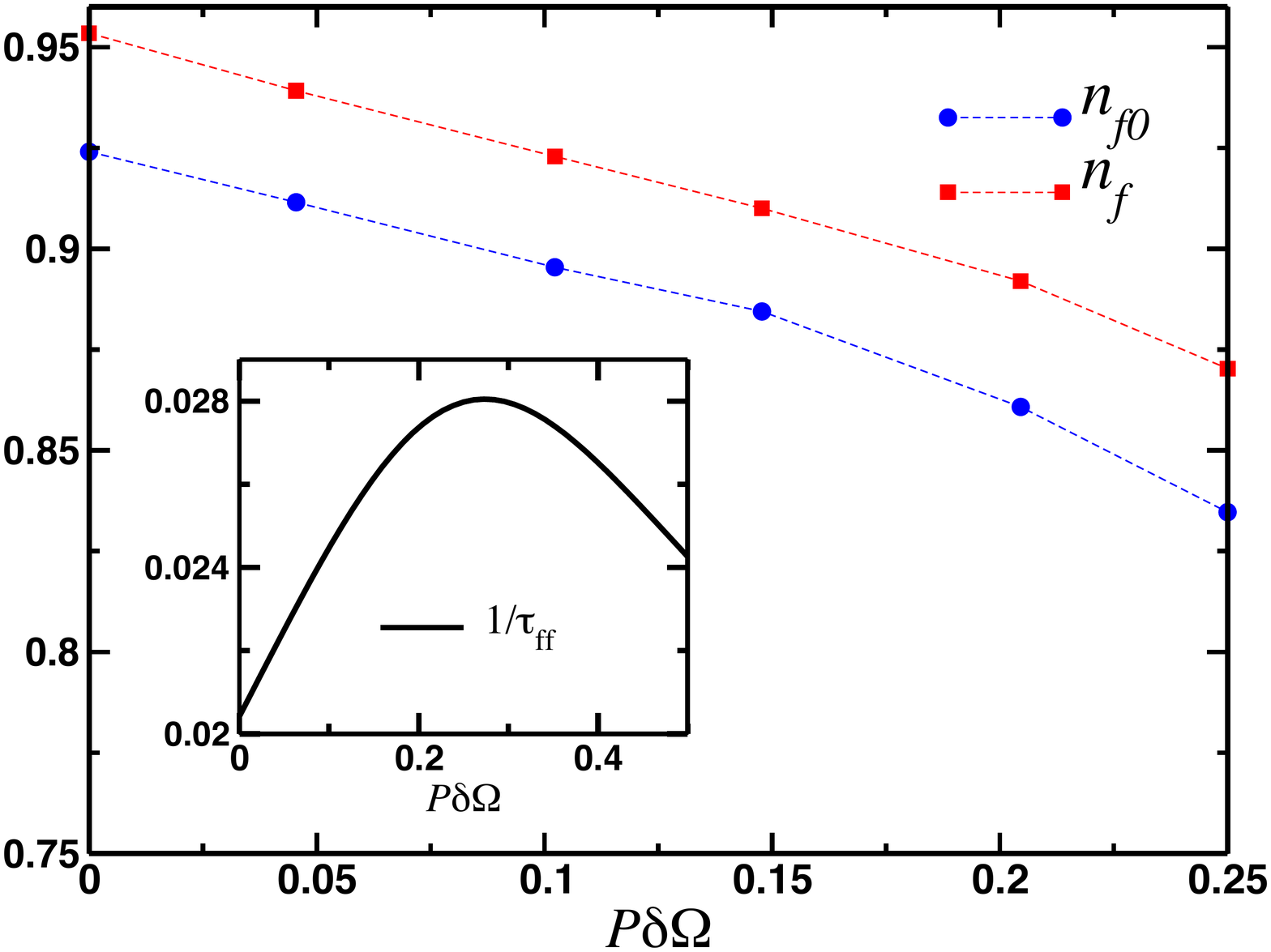}
\caption{\small(Color online) Plot of the $f$-electron occupation number at the mean-field level $n_{f0}$ and the occupation
number beyond mean-field $n_f$ and a function of tensile strain $P\Delta\Omega>0$. The inset shows the relaxation time for the 
$f$-electrons due to fluctuations $1/\tau_{ff}=\textrm{Im}\Sigma_{ff}(i\omega_0)$ where the imaginary part
of the self-energy is computed at zero Matsubara frequency $\omega_0=\pi T$. All energies are presented in the units of the 
half width of the conduction band.}
\label{SM-TheoryFig}
\end{figure}
To find the correction to the $f$-electron propagator
\beg\label{Gf}
G_{ff}(k,k')=-\langle\hat{T}_\tau f_m(\bk,\tau)\overline{f}(\bk',\tau')\rangle
\en
we take into account that the fluctuating field $i\lambda$ couples to the $f$-electron density. Therefore, the the self-energy correction
to the $f$-electron propagator due to fluctuations of the constraint field obtains
\beg\label{Sigmaff}
\Sigma_{ff}(\bk,i\omega)=T\sum\limits_{i\nu}\sum\limits_{\bp}D_\lambda(\bp,i\nu)G_0^f(\bk+\bp,i\omega+i\nu).
\en
At very low temperatures $T\to 0$ and small Matsubara frequencies $i\nu$, $D_\lambda(\bp,i\nu)$ shows weak momentum dependence apart from the very narrow region in the some small values of momenta, where it has a small peak with the width $\approx T$. Therefore, we can approximately write for the self-energy the following expression
\beg\label{AppSigmaff}
\Sigma_{ff}(i\omega)\approx T\sum\limits_{i\nu}D_\lambda(i\nu)\sum\limits_{\bp}G_0^f(\bp,i\omega+i\nu),
\en
which implies that the self-energy becomes purely local. Thus, for (\ref{Gf}) we write
\beg
G_{ff}(\bp,i\omega)=\frac{G_0^f(\bp,i\omega)}{1-G_0^f(\bp,i\omega)\Sigma_{ff}(i\omega)}.
\en
Then, equation which determines the particle number reads
\beg
n_f=T\sum\limits_{i\omega_n}e^{i\omega_n0+}\sum\limits_{\bp}G_f(\bp,i\omega)
\en
The results of the calculation are shown in Figure \ref{SM-TheoryFig}.

Note that the fluctuations tend to slightly increase the $f$-electron occupation number favoring the magnetic valence configuration for small to moderate values of $P\delta\Omega$ while the imaginary part of the self-energy grows with an increase in $P\delta\Omega$. 
However, with further increase in $P\delta\Omega$ the imaginary part reaches maximum and then decreases. We also find that the correction to the $f$-electron number decreases as well. This observation confirms our expectations that for large enough values of the strain field, the time scale corresponding to the change in the intervalence configuration should increase. Importantly, our results indicate that at ambient pressure the fluctuations in Kondo insulators must play an important role in determining the thermodynamic properties of these systems, while the application of the tensile strains renders systems to be a more 'mean-field' like.

\bibliography{strainsmb6}

\end{document}